\documentclass[reqno]{amsart}
\usepackage{graphicx,amsmath}

\newcommand{\pd}[2]{\frac{\partial #1}{\partial #2}}

\newcount\n  \newdimen\w

\def\Repeat#1#2{\n=#1\relax\loop\ifnum       
	\n>0\relax #2\advance\n by-1\repeat}

\long\def\OMIT#1{\relax }  

\def\re#1{(\ref{#1})}   

	\def\delim#1#2#3{\csname\ifcase#1 relax\or   
		big\or Big\or bigg\or Bigg\fi\endcsname   
		{\ifcase#2\or\Delim#3\or\deliM#3\fi}}      
	\def\Delim#1{\ifcase#1\relax\or(\or[\or\{\or<\or\langle\or|\or\|\or---{ }\fi}
	\def\deliM#1{\ifcase#1\relax\or)\or]\or\}\or>\or\rangle\or|\or\|\or{ }---\fi}
	
	\def\largerfrac#1#2#3{      
		\whichtypesize\n=\currenttypesize\advance\n by #1 \mathchoice
		{\setbox0\hbox{$\displaystyle-$} \w=.5\ht0\advance\w by-.5\dp0\setbox0
			\hbox{\typesize\n $\displaystyle-$} \advance\w by -.5\ht0\advance\w
			by .5\dp0\raise\w \hbox{\typesize\n$\displaystyle{\frac{#2}{#3}}$}}
		{\setbox0\hbox{$-$} \w=.5\ht0 \advance\w by -.5\dp0 \setbox0\hbox
			{\typesize\n $-$} \advance\w by-.5\ht0\advance\w by
			.5\dp0\raise\w\hbox{\typesize\n$\frac{#2}{#3}$}}
		{\setbox0\hbox{$\scriptstyle-$} \w=.5\ht0 \advance\w by-.5\dp0\setbox0
			\hbox{\typesize\n $\scriptstyle-$} \advance\w by -.5\ht0 \advance\w
			by .5\dp0 \raise\w\hbox{\typesize\n$\scriptstyle{\frac{#2}{#3}}$}}
		{\setbox0\hbox{$\scriptscriptstyle-$} \w=.5\ht0
			\advance\w by -.5\dp0 \setbox0\hbox{\typesize\n
				$\scriptscriptstyle-$} \advance\w by -.5\ht0 \advance\w by .5\dp0
			\raise\w\hbox{\typesize\n$\scriptscriptstyle{\frac{#2}{#3}}$}}  }

	       \def\pd{\partial}

	\newcommand{\be}{\begin{equation}}
		\newcommand{\ee}{\end{equation}}
	\newcommand{\ben}{\begin{equation*}}
		\newcommand{\een}{\end{equation*}}
	\newcommand{\bea}{\begin{eqnarray}}
		\newcommand{\eean}{\end{eqnarray*}}
	\newcommand{\bean}{\begin{eqnarray*}}
		\newcommand{\eea}{\end{eqnarray}}

	\newcommand{\todo}[1]{}
	\renewcommand{\todo}[1]{\textcolor{blue}{\small\texttt{[:~#1~:]}}}

\begin{document}
		
		\title{About the dissipative Newton equation}
		\author{P. V\'an$^{1,2,3}$ }
		\address{$^1$Department of Theoretical Physics, HUN-REN Wigner Research Centre for Physics, H-1525 Budapest, Konkoly Thege Mikl\'os u. 29-33., Hungary;\\
			and  $^2$Department of Energy Engineering, Faculty of Mechanical Engineering,  Budapest University of Technology and Economics, H-1111 Budapest, M\H{u}egyetem rkp. 3., Hungary;\\
		$^3$Montavid Thermodynamic Research Group, Hungary}
		\date{\today}
		
		\begin{abstract}
			The thermodynamic basis of classical mechanics is presented. In this framework, ideal Newtonian mechanics emerges as the zero-dissipation limit of a more general, dissipative theory. The thermodynamic approach predicts a novel dissipative contribution to the momentum that depends on the applied force, leading to a damping coefficient with a specific, experimentally testable dependence on the inertial mass and the spring constant. A torsion balance experiment with variable moment of inertia has been designed to measure this effect. Several known equations, including a thermodynamic version of the Eliezer-Ford-O'Connell equation of radiation reaction, are recovered as special cases.
		\end{abstract}
		\maketitle

\section{Introduction}

	What is dissipation? The only clear answer can be formulated within a thermodynamic framework that defines whether entropy is produced during the evolution of a physical system. A thermodynamic framework separates external entropy transport from internal production and clarifies whether and how the evolution equations -- field equations and equations of motion -- of the system contribute to entropy production. Such an embedding requires that the basic physical quantities, the variables that determine the evolution of the system, are thermodynamic state variables at the same time. Then, and only then, entropy production is the only truly fundamental measure of dissipation. 
	
	However, only a few theories are thermodynamically compatible; most of them are ideal, based on variational principles, and dissipation appears only as a particular and practical extension of their evolution equations. One can distinguish three types of extensions: ideal, dissipative, and impossible. Impossible modifications are mostly unphysical and frequently lead to unstable initial value problems. It is common to connect dissipative extensions to stability. There is a relation of thermodynamic compatibility and stability \cite{Mat05b,Had19b}, but it is not straightforward \cite{SomEta25a,GioZul26a}.  
	
		
	In any case, thermodynamic compatibility is a conservative method to define dissipative processes and find viable physical extensions of evolution equations. It is also remarkable that checking the compatibility of already existing evolution equations is not the only possibility: the thermodynamic framework can be used constructively: evolution equations can be derived. The thermodynamic methodology was recognised in Extended Thermodynamics \cite{Gya77a,MulRug98b,JouAta92b}. There, compatibility with kinetic theory is a method of validation. 	
		
The methodology of constructing evolution equations has been tested far beyond the validity of microscopic theory. The extension of Extended Thermodynamics is the method of internal variables. Those are introduced for the characterisation of complex materials with complicated interactions, where drawing conclusions from detailed microdynamics is hopelessly complicated \cite{MauMus94a1,MauMus94a2}. In such cases, there is no way to obtain their evolution equations from micro- or mesodynamics; only general principles are applicable and the tests are the experiments and also the compatibility with extended thermodynamics \cite{KovEta21a}. Without a microscopic backround there are two seemingly incompatible general approaches: thermodynamical evolution with relaxation and mechanical evolution characterised by inertia. In the case of internal variables, both evolution types are observable and important. 
	
	Let us remark: relaxational evolution seems to be purely dissipative, while evolution with inertial effects always has an ideal core without dissipation. For example a point mass keeps moving forever in an ideal world, in the absence of dissipative effects. The unification of the two approaches is not evident. Mechanics is built on Hamiltonian variational principles, and those principles are not applicable to dissipative processes without further ado \cite{VanNyi99a}. It is not impossible to construct variational principles for dissipative processes, but the required modifications are neither natural nor general; moreover, it is even more problematic to find a variational principle that applies to both ideal and dissipative evolution. Therefore, it is common to double the theoretical framework and merge the variational principles of ideal mechanics with thermodynamic requirements, as is done for example in GENERIC \cite{GrmOtt97a,OttGrm97a,Ott05b}. Mechanical concepts are not really suitable for dissipative evolution.  
	
	However, the other way around, the thermodynamic method can be easily generalised for systems with inertia: both dissipative and ideal evolution can be constructed from the Second Law of Thermodynamics in one stroke. The idea is to represent the inertial effect by a second internal variable in the method of dual internal variables \cite{VanAta08a,BerVan17b}. Here, ideal mechanical evolution is the marginal case of the dissipative one, with zero dissipation. From a thermodynamic point of view, one may wonder whether ideal mechanics is the marginal theory of a more general dissipative one. 
	
	At first glance it seems straightforward to introduce position and momentum as thermodynamic state variables and to create the particular dissipative analogy with the Hamiltonian equations. At a second glance, there are several problems with this fundamental Aristotelian step \cite{MarRop87a}. Are position and momentum suitable thermodynamic state variables? Are the kinematics, spacetime embedding, and Galilean covariance compatible with thermodynamics? In any case, the ideal equations are the same at the end. 
	
	In this paper I look for practical, experimentally testable consequences arguing that first-principle dissipation may lead to remarkable differences that are not evident from the conceptual framework of ideal mechanics and could be tested experimentally. First, the simplest dissipative extension of the Newton equation is derived by analogy with dual internal variables. Then I show that the momentum is no longer proportional to velocity; it may have dissipative parts. Finally, I argue that this qualitative difference leads to a particular damping coefficient that cannot be predicted by mechanics alone and could be measured with properly designed experiments. I also discuss other special cases and some interpretations and other possibilities to validate the prediction.

\section{The logic of mechanics and thermodynamics}

There are two different approaches to finding evolution equations for internal variables in non-equilibrium thermodynamics. One assumes that inertia is the primary effect; the other assumes that internal variables are relaxational. Actually, it is not an assumption, it is a presumption, wired in the mathematical framework. The observed evolution is usually something between the two; therefore, both mechanisms must be considered. A unification is problematic, as the two approaches are seemingly incompatible. Naturally, one may mix and combine the two approaches as in the case of phase fields \cite{FabEta06a}, also taking into account their compatibility \cite{Ott05b}, but that requires independent postulates and mathematical structures for both the inertial and relaxational parts. However, inertia is based on momentum, and this simple observation can unify the two structures, as was suggested by the method of dual internal variables \cite{VanAta08a,BerVan17b}. Then, the double set of principles is avoided and one can obtain ideal evolution in the limit of zero dissipation. The Second Law of Thermodynamics becomes the primary principle behind any internal variable evolution.

\section{Thermo-Dynamics of two variables}

In the case of dual internal variables, the entropy function is the generator of the evolution. For the sake of easier concept matching, let us assume that the total energy of the mechanical system is conserved and the internal energy is the difference of the total energy and the Hamiltonian, therefore $S(E,x,p) = \hat S(U=E-H(x,p))$. Mechanical notation and terminology is used within the internal variable framework of non-equilibrium thermodynamics.

Now we are looking for the dynamics of the system of two physical quantities $x$ and $p$ by the following dynamics:
\bea
\dot{x}&=&g_x(x,p), \label{gx}\\
\dot{p}&=&g_p(x,p). \label{gp}
\eea

Here the right hand side is to be determined by the requirement of thermodynamic compatibility. That is, first of all, according to thermodynamic stability, entropy is a concave function, and then the Hamiltonian $H$ is convex. Moreover, the entropy production rate is nonnegative; entropy is increasing along the differential equation, therefore
\be
\dot{S}(E,x,p) =\pd_E S(-\pd_xH)\dot x + \pd_E S(-\pd_pH)\dot p = -\pd_E S\left(\pd_xH g_x + \pd_pH g_p\right) \geq 0
\label{SL_therm}\ee
where we have introduced the notation $\pd_x S = \left.\frac{\pd S}{\pd x}\right|_{E,p}$, and analogously for the other partial derivatives of $S$ and $H$. It was also assumed that the total energy $E$ is constant, and $\pd_E S = \frac{d \hat S}{d U} =: \frac{1}{T}$ is the reciprocal temperature in this framework. Complex friction models are examples of open discrete thermomechanical systems where the energy is not conserved \cite{MitVan14a}.  

Therefore one obtains that the Hamiltonian along the differential equation must decrease:
\be
\dot{H}(x,p) = \partial_xH g_x(x,p) + \partial_pH g_p(x,p) \leq 0
\label{SL_mech}\ee

According to the constructive thermodynamic approach, we are looking for such an evolution equation \re{gx}-\re{gp} that does not violate the above inequality. The general solution of \re{SL_mech} determines the unknown functions $g_x$ and $g_p$ according to the Lagrange mean value theorem as:
\be
\begin{pmatrix} 	\dot x \\\dot p \end{pmatrix} =-
\begin{pmatrix} l_{1} & l_{12} \\ l_{21} & l_{2} \end{pmatrix}
\begin{pmatrix} \pd_x H \\\pd_p H \end{pmatrix}, 
\ee
where the symmetric part of the matrix $L$ is positive definite.  It is convenient to separate its symmetric and antisymmetric parts as follows
\be
L=\begin{pmatrix} l_{1} & l_{12} \\ l_{21} & l_{2} \end{pmatrix}=
\begin{pmatrix} l_{1} & l \\ l & l_{2} \end{pmatrix}+
\begin{pmatrix} 0 & k \\ -k & 0 \end{pmatrix}.
\ee
In general, the coefficients are functions of the state space, but in the following they are considered as constants: this is the  linear approximation of the general solution. 

The inequality \re{SL_mech} is valid for any $g_x$ and $g_p$, therefore the $L$ matrix is not arbitrary, the following inequalities hold for the coefficients:
\be
l_1 \geq 0, \qquad
l_2 \geq 0, \qquad
l_1l_2-l^2  \geq 0.
\label{SL_ineq}\ee
Also, the determinant of $L$, $\det L = l_1l_2-l^2 +k^2 \geq 0$, follows as a consequence. The symmetric part of the matrix is responsible for the dissipative part of the dynamics. If it is zero, then the inequality \re{SL_mech} is fulfilled as an equality, the Hamiltonian is conserved, and the entropy is constant. Separating the ideal and dissipative parts of the dynamics, the following differential equations emerge:
\bea \dot{x} &=& k \partial_pH - l_1 \partial_xH - l \partial_pH
=k \partial_pH + f_x, \label{td1}\\
\dot{p} &=& -k \partial_xH - l \partial_xH - l_2 \partial_pH
=  -k \partial_xH +f_p,\label{td2}
\eea

Here the dissipative part of the dynamics is denoted by $f_x$ and $f_p$. The antisymmetric part defines a symplectic Hamiltonian evolution equation if $k=1$. The dimensionless parameter $k$ matches the kinematics to the dynamics in ideal mechanics. In our thermodynamic framework we keep $k$ as it is, to make the dissipative and ideal parts of the evolution equation easily distinguishable.

\subsection{Example of conservative mechanics}

The previous consideration is a simplified but general scheme of dual internal variables, but with the terminology of mechanics and minimalising the thermodynamical framework. However, one can identify some interesting and unavoidable consequences for pure mechanics, too. Therefore, let us further specify the theory, and rewrite the above system with the following Hamiltonian:
\be
H(x,p) = \frac{p^2}{2m} + V(x).
\label{Hf} \ee

The physical interpretation is straightforward according to mechanical principles. $V$ is the ``mechanical'' potential, $m$ is the mass, and $p$ is the momentum. Therefore \re{td1} and \re{td2} are:
\bea 
\dot{x} &=& \frac{k}{m} p - l_1 V' - \frac{l}{m} p,\label{tdc1}\\
\dot{p} &=& -k V' - l V' - \frac{l_2}{m} p.\label{tdc2}
\eea

Here $V'$ is the derivative of the mechanical potential and $F:= -V'$ is the force. One can get the "momentum" $p$ from \re{tdc1} as
\be
p(\dot{x},x)=m\frac{\dot{x}+l_1 V'}{k-l} \label{mom}.
\ee
Eliminating the momentum $p$ results in the following equation of motion:
\be
m\ddot{x} + (m l_1 V'' + l_2)\dot{x} + (l_1 l_2- l^2 + k^2 )V' = 0.
\label{em}\ee

The equation looks like a normal equation of motion of, a Newton equation with damping. However, there are differences and extra consequences. The coefficients are nonnegative if $V$ is convex (due to thermodynamic compatibility) and due to the nonnegative entropy production. Therefore, the equilibrium of the equation is asymptotically stable. Comparison with Lagrangian dynamics reveals that the second term is responsible for the damping; however, its particular form differs from our expectations: the coefficient of $\dot{x}$ depends on $V$ and also on the mass. In mechanics, a typical damping term should not depend on the primary characteristic parameters of the motion, like the spring constant. The last term, normally considered as a result of the nondissipative part, contains the dissipative coefficients $l$, $l_2$, $l_1$; the force becomes enhanced due to dissipation. However, the difference is quantitative, and minor differences cannot be recognised easily from a purely mechanical point of view. In a mechanical interpretation through Hamiltonian dynamics with damping, one can easily mix the dissipative terms with the nondissipative ones.

Let us observe that in the special case of $f_x=0$ (that is, $l=0$ and $l_1=0$), the distinction between the dissipative and non-dissipative parts is straightforward: the second term represents the classical damping, which is negatively proportional to the velocity. The unexpected new term is originated from upper left corner of the matrix $L$, which modifies the definition of the momentum: if $l_1 \neq 0$, then it is no longer proportional to the velocity alone.

\section{Dissipation and Hamiltonian mechanics}

One can try to separate the dissipative and nondissipative parts of the above dynamics in a different way. Let us consider \re{td1} and \re{td2} and formulate a Hamiltonian reconstruction. This can be done with the following Lagrangian:
\be
L(\dot{x},x) = -H(x,p) + \frac{1}{k}\left(p\dot{x} -\int f_x dp
\right).
\ee

Here the Lagrangian is given through a Legendre transformation as in the case of pure mechanics. Therefore (compare to \re{AH}--\re{CH} of the Appendix): 
\bea
\partial_{\dot{x}}L=\frac{p}{k}, \label{dAH}\\
\partial_{p}H + \frac{f_x}{k}=\frac{\dot{x}}{k}, \label{dBH}\\
\partial_{x}L=-\partial_{x}H - \frac{1}{k}\int \partial_xf_x dp, \label{dCH}
\eea
where \re{dBH} defines $p(\dot{x},x)$ and corresponds exactly to \re{td1}.
\re{dAH} and \re{dCH} can be combined to get an Euler-Lagrange
equation as
\bea
\frac{d}{dt}\partial_{\dot{x}}L &-& \partial_{x}L = \frac{\dot{p}}{k}
+ \partial_{x}H + \frac{1}{k}\int \partial_xf_x dp \nonumber\\
&=& \frac{1}{k}\left(\int \partial_xf_x dp + f_p\right) = F_{diss}.
\label{dELH}\eea

Here in the second equality we have used \re{td2}, and contrary to \re{ELH}, the right hand side of the equation does not vanish; it is the dissipative contribution, denoted by $DiF$. This is how the dissipative part of mechanics is identified by the Hamiltonian structure.

\subsection{Conservative-like dissipative mechanics}

Let us see again the simple Hamiltonian \re{Hf} defined above. In this case the dissipative Lagrangian is:

\be
L(\dot{x},x) = -H(x,p) + \frac{1}{k}\left(p\dot{x} + l_1 V' p +
\frac{l}{2m} p^2\right)
\ee

Calculating the equations \re{dAH}-\re{dCH} one can get the momentum as \re{mom} and therefore the Lagrangian is
\be
L(\dot{x},x) = \frac{m}{2 k(k-l)} (\dot{x} + l_1 V'(x))^2 - V(x).
\ee

The corresponding Euler-Lagrange equation follows as
\bea
\frac{d}{dt}\partial_{\dot{x}}L &-& \partial_{x}L =
\frac{m}{k(k-l)}\left(\ddot{x} - (l_1^2 V''
+\frac{k^2-kl}{m})V'\right) = 0 \label{dcELH}\eea

The additional "dissipative" part, denoted by $F_{diss}$ in \re{dELH}, is
\bea
F_{diss} = \frac{m}{k(k-l)}\left((\frac{l_2}{m} + l_1 V'')\dot{x} + (l_1^2
V'' +\frac{kl-l^2+l_1l_2}{m})V'\right) \label{ddpart}
\eea

Adding \re{dcELH} and \re{ddpart} together gives \re{em}, in accordance with our expectations. However, we can observe that the ideal and dissipative elements of the structure are mixed when compared with the initial entropy production based identification. It is also remarkable, that in general $DiF$ cannot be derived from a dissipation potential.


\section{Special cases: the role of force}

It is instructive to rewrite the system \re{tdc1}-\re{tdc2} by substituting the negative derivative of the potential, the force $F= -V'$:
\bea 
m \dot{x} &=& l_1 m F + (k-l) p,\label{tdf1}\\
\dot{p} &=& (k +l)F - \frac{l_2}{m} p.\label{tdf2}
\eea

Eliminating the momentum one obtains:
\be
m\ddot x - m l_1 \dot F +l_2 \dot x - {\rm det}L\,\, F = 0. \label{mfeq_gen}
\ee
Here ${\rm det}L = l_1l_2-l^2 +k^2$ is the determinant of $L$. It is remarkable that the effective coefficients of \re{mfeq_gen} are interdependent; only the original thermodynamic coefficients can be chosen independently, otherwise the stability of the equilibrium can be easily destroyed \cite{SomEta25a}.

\subsection{Newtonian mechanics: $l_1 = 0$ and $l_2 = 0$}
Then the third inequality of \re{SL_ineq} requires that $l=0$, too. The system of equations \re{tdf1}-\re{tdf2} reveals the structure of Hamiltonian mechanics:
\bea 
m \dot{x} &=& k p,\label{h1}\\
\dot{p} &=& k F,  \label{h2}
\eea
\noindent and the Newtonian equation of motion emerges in particular if the scale factor $k=1$:
\be
m\ddot x  = F. \label{Neq}
\ee

\subsection{Damped mechanics:  $l_1 = 0$}

Then according to the third inequality of \re{SL_ineq}, $l=0$ again. The system of equations \re{tdf1}-\re{tdf2} reduces to the structure of damped Hamiltonian mechanics:
\bea 
m \dot{x} &=& k p,\label{dh1}\\
\dot{p} &=& k F - \frac{l_2}{m} p.\label{dh2}
\eea

Eliminating the momentum and setting $k=1$, one obtains:
\be
m\ddot x + l_2 \dot x - F = 0. \label{dNeq}
\ee
The equation is overdamped if the inertial term is negligible.

\subsection{Eliezer-Ford-O'Connell-Verh\'as equation: $l_2 = 0$}
Then according to \re{SL_ineq}, $l=0$ as in the previous cases. The system of equations \re{tdf1}-\re{tdf2} reduces to the following form:
\bea 
m \dot{x} &=& l_1 m F + k p,\label{radHeq1}\\
\dot{p} &=& k F .\label{radHeq2}
\eea

Eliminating the momentum and setting $k=1$ as before, one obtains:
\be
m\ddot x - m l_1 \dot F - F = 0. \label{EFOVeq}
\ee

This is the thermodynamic version of the Eliezer-Ford-O'Connell equation of radiation reaction, which is expected to eliminate runaway solutions \cite{Eli43a,ForOCo91a,PriEta22a}. 

\subsection{Aristotelian resonance: $l=k$}

Finally, let us regroup the terms of \re{mfeq_gen} as:
\be
m(\ddot x - l_1 \dot F) +l_2 \left(\dot x - \frac{{\rm det}L}{l_2} F \right)= 0. \label{Aheq}
\ee
This is the sum of an overdamped relaxational equations and the time derivative of another one with a different damping coefficient. Putting them together, the solutions may not be relaxational at all, due to the presence of the second order time derivative. One may expect that relaxational dynamics can appear only if the effect of the first term is negligible, e.g.\ if $m=0$. However, if $k=l$, the two terms have the same structure and one can obtain overdamped relaxational dynamics for large masses as well. This hierarchical structure is similar to the Fourier resonance of the Guyer-Krumhansl equation \cite{VanEta17a1}.

\section{Dissipative momentum: experimental prediction}

The previous examples demonstrate that a thermodynamic framework reveals some unexpected and original aspects of Newtonian mechanics. In particular, the velocity-momentum relation, equation \re{mom}, have a dissipative and force-dependent part.

Testing that property experimentally would be a strong argument for the thermodynamic approach. Consider, for example, a damped harmonic oscillator with $V(x) = D x^2/2$, where $D$ is the spring constant. Then equation \re{em} takes the following simple form:
\be
m\ddot{x} + (m l_1 D + l_2)\dot{x} + {\rm det} L Dx = 0.
\label{oem}\ee

Moreover, a further reparametrisation with the effective spring constant $\hat D = (l_1 l_2- l^2 + k^2 ) D$, $a:= \frac{l_1}{{\rm det} L}$, and $b=l_2$ yields
\be
m\ddot{x} + (m \hat D a + b)\dot{x} + \hat D x = 0.
\label{opem}\ee
Here the linear dependence of the damping coefficient $\hat b= (m \hat D a + b)$ on the inertial mass and on the spring constant looks like a universal and experimentally measurable prediction of the thermodynamic framework. 

\begin{figure}
	\begin{center}
	\includegraphics[width=.5\textwidth]{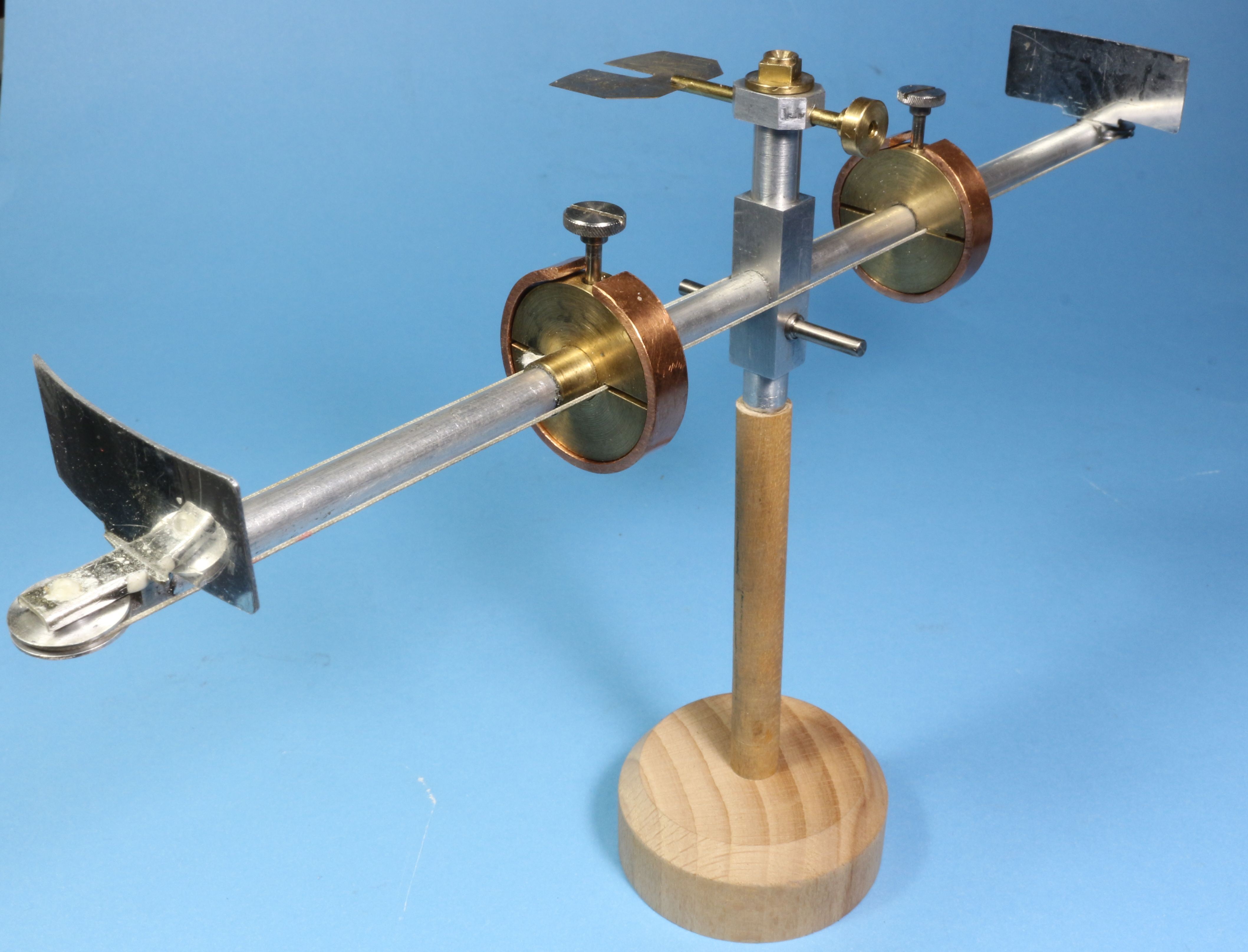}
	\end{center}
	\caption{ \label{ingatest} The arm of the torsion balance. The upper part at the centre is related to the optical readout. The wires ensure the parallel movability of the masses. When operational, it is suspended upside-down.}
\end{figure}

It may be a small effect under normal circumstances and therefore requires a sensitive experimental device, whose main aspect is that the inertia is variable. We have constructed such an experimental device, based on a Bessel-type torsion balance, where the moment of inertia is variable due to the possibility that the test masses are mobile along the arms of the balance. In Figure \ref{ingatest} one can see the arm with the synchronously movable dials.
 
 The whole instrument is shown on Figure 2. The torsion wire is in the upper cylindrical part and the lower part is the vacuum chamber of the arm. Below is the box of the control unit.
	\begin{figure}
	\begin{center}
	\includegraphics[width=.6\textwidth]{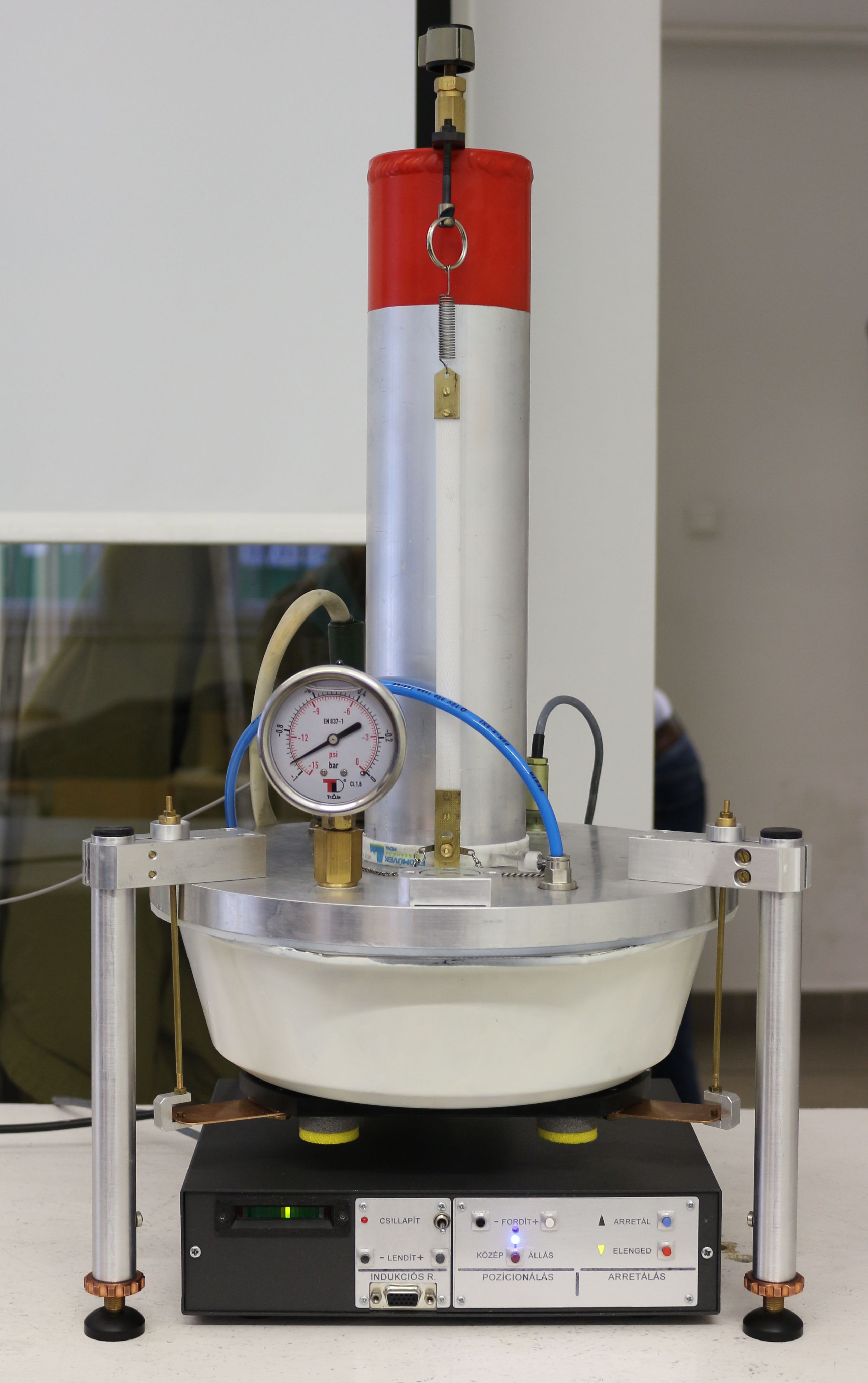}
	\end{center}
	\caption{ The torsion balance with variable moment of inertia in its vacuum chamber. Below is the electronic control unit.}
\end{figure}

\section{Discussion}

The central result of this paper is the thermodynamic embedding of Newtonian mechanics. The ideal Newton equation of motion emerges as the limit of zero dissipation from a framework governed by the Second Law of Thermodynamics. The $l_1$ coefficient in the momentum relation \re{mom} is an original consequence of the thermodynamic approach: a dissipative, force-dependent part in the momentum-velocity relation. This term has no counterpart in standard Hamiltonian mechanics and leads to an experimentally testable prediction: the damping coefficient depends linearly on both the inertial mass and the spring constant, as it is shown in   \re{opem}. 

Given a dissipative structure, it is straightforward to identify the Hamiltonian, non-dissipative part. Given a Hamiltonian structure, however, it is far from obvious how to add correct dissipation terms. As we have seen, a purely mechanical decomposition easily mixes the ideal and dissipative contributions, whereas the thermodynamic framework keeps them separated from the outset.

\subsection{Micro-macro relation of dissipation}

There is a broad consensus concerning the microscopic origin of dissipation. It is widely accepted that fundamental theories are ideal, and dissipation is interpreted as an effective phenomenon attributed to averaged or coarse-grained microdynamics. Despite the evident mathematical and conceptual difficulties \cite{Mac92b}, the macroscopic dissipative parameters can be calculated if the  microscopical constituents and their dynamics is known. However, one cannot explain dissipation solely with ideal microscopic concepts, because one must first have a concept of dissipation itself. How can we know that a macroscopic evolution equation is truly dissipative, without a theory that defines the concept? Which parameter must have a definite sign according to the Second Law?

The particular nature and mechanisms of microscopic dissipation may be manifold. Kinetic theory is the most important candidate for microscopic explanation, and since Boltzmann there are several ideas beyond the ``Stosszahlansatz'', the assumption of microscopic reversibility. From a practical point of view, the usefulness of such explanations is limited, because the assumed microscopic structure, like colliding particles, remains special.

On the other hand, according to our experience, the Second Law for macroscopic physics is universally valid, independently of the particular microstructure. The universality of the temperature, for instance, follows from the universal efficiency of the Carnot process \cite{Tho849a,Van23a1}. Therefore, an everlasting important subject of basic thermodynamic research is to understand macroscopic physics independently of microscopic dynamics and structures. The separation of the universal part of macrophysics -- in particular the Second Law and the related thermodynamic concepts -- is the fundamental vision of non-equilibrium thermodynamics \cite{Van20a}.

The present work applies nonequilibrium thermodynamics to the simplest mechanical setting. The dissipative extension of the Newton equation is derived from the Second Law alone, without any assumption about microscopic mechanisms. The resulting structure contains both the familiar velocity-proportional damping (controlled by $l_2$) and the novel force-dependent momentum correction (controlled by $l_1$). While the former is routinely attributed to collisional or viscous microscopic processes, the latter has no standard microscopic explanation. It is a genuinely macroscopic, thermodynamic prediction. Various microscopic mechanisms may underlie the $l_1$ term -- delay effects, memory in the interaction, or non-local coupling -- but the thermodynamic framework does not require their specification and remains valid regardless of which mechanism dominates.

\subsection{Outlook}

The structure of Eliezer-Ford-O'Connell equation of radiation reaction \re{EFOVeq}, recovered here as the $l_2=0$ special case, illustrates that the thermodynamic framework naturally encompasses phenomena usually derived from electrodynamics. The thermodynamic background of this connection as inertial effect was first analysed by Verh\'as \cite{Ver89a}. 

Regarding possible microscopic origin of dissipative momentum a physical interpretation of the Ehrenfest regularisation and the lack-of-fit reduction can give ideas \cite{MlaEta25m,PavEta19a}.

At the time of the first presentation of this work, only the experimental device was available. Since then, measurements have been started. The torsion balance with variable moment of inertia is designed to detect the mass- and spring-constant-dependent damping predicted by equation \re{opem}. 

There may be further consequences of the thermodynamic embedding. A non-local extension coupled to continuum theory could be a natural next step, connecting the present discrete framework to field-theoretic descriptions. Applications to rheology, friction models \cite{MitVan14a}, and the Lorentz-Abraham-Dirac equation of radiation reaction are also worth exploring.

There are also several aspects where the presented framework is oversimplified. For example, the step from the scalar thermodynamic internal variable representation of position toward a three dimensional Euclidean space is straightforward, but it is far from a proper Galilean covariant theory. Also, the dissipative, generalised velocity-momentum relation should be interpreted with the geometrical concepts of mechanics.

\section{Acknowledgement}

In memory of my late mentor, Joe Verhás, from whom this insight originated.

The author acknowledges networking support by the grant NKFIH NKKP-Advanced 150038 and COST Action FuSe CA24101. 

The research reported in this paper is part of project no. TKP-6-6/PALY-2021, implemented with the support provided by the Ministry of Culture and Innovation of Hungary from the National Research, Development and Innovation Fund, financed under the TKP2021-NVA funding scheme.

\section{Appendix: Pure mechanics}

Hamiltonian mechanics is generated by a Hamiltonian, $H(x,p)$. For a single point mass at the position $x$ and with the momentum $p$, the Hamilton equations are: 
\bea
\dot{x} &=& \partial_p H, \label{H1}\\
\dot{p} &=& -\partial_x H, \label{H2}
\eea
where the dot denotes the time derivative and $\partial_p$ and $\partial_x$ are the partial derivatives of the Hamiltonian with respect to the momentum and the position, respectively.
The Lagrangian $L(x,\dot x)$ is obtained by Legendre transformation:
\be	
L(x,\dot{x}) + H(x,p) = p \dot{x} \label{Lt}.
\ee

Therefore
\bea
\partial_{\dot{x}}L=p, \label{AH}\\
\partial_{p}H=\dot{x}, \label{BH}\\
\partial_{x}L=-\partial_{x}H. \label{CH}
\eea

The Euler-Lagrange equation is obtained by eliminating the momentum $p$:
\be
\frac{d}{dt}\partial_{\dot{x}}L- \partial_{x}L =
\dot{p} + \partial_{x}H=0.
\label{ELH}\ee

Let us observe the structure of the equations. \re{H1} is \re{BH} and is exploited to get $p(\dot{x},x)$. At the first equality of \re{ELH} we substituted \re{AH} and \re{CH}, and in the last equality \re{H2} was used. Let us remember: the Hamiltonian is the primary concept and we have defined the Lagrangian through a Legendre transformation in \re{Lt}. In Hamiltonian mechanics equations \re{AH}--\re{ELH} are consequences \cite{Arn78b}.

\bibliographystyle{unsrt}

\end{document}